\documentclass[review]{elsarticle}

\usepackage{lineno,hyperref}
\modulolinenumbers[5]

\begin{document}

\begin{frontmatter}

\title{To the synthesis and characterization of layered metal phosphorus triselenides proposed for electrochemical sensing and energy applications (Manuscript was initially submitted to ACS Catalysis as a Comment for Ref. 22; however was rejected after appeal)}

\author[SHU,SFedU]{Yuriy Dedkov\corref{corr1}}
\cortext[corr1]{Corresponding author}
\ead{dedkov@shu.edu.cn}

\author[SHU]{Mouhui Yan}

\author[SHU,SFedU]{Elena Voloshina\corref{corr2}}
\cortext[corr2]{Corresponding author}
\ead{voloshina@shu.edu.cn}

\address[SHU]{Department of Physics, Shanghai University, 200444 Shanghai, China}
\address[SFedU]{Institute of Physical and Organic Chemistry, Southern Federal University,\\ 344090 Rostov on Don, Russia}

\begin{abstract}

Recent studies reported on the synthesis and characterization of several bulk crystals of layered metal triselenophosphites MPSe$_3$ (M = transition metals). In these works characterization was performed via a combination of different bulk- and surface-sensitive experimental methods accompanied by DFT calculations. However, the critical examination of the available experimental and theoretical data demonstrates that these results do not support the conclusions on the electrochemical sensing and energy applications of studied triselenophosphites. These conclusions are made without any relation to the age of discussed data and possible recent progress in experimental and theoretical approaches.

\end{abstract}


\end{frontmatter}


\section{Introduction}

The discovery of the unique transport properties of graphene in in 2004~\cite{Novoselov:2005es,Zhang:2005gp} lead to the increased attention to graphene-based systems~\cite{Dedkov:2015kp,Janthon:2013cj,Roy:2017fg,Dedkov:2020da} and also to other classes of 2D materials. Among them are h-BN~\cite{Oshima:1997ek}, two-dimensional dichalcogenides~\cite{Manzeli:2017ib,Meng:2020asd}, and many others, as well as heterosystems on the basis of these materials~\cite{Geim:2014hf}. These studies led to the discovery of many exciting properties, which can be used in different applications, like, touch screens~\cite{Bae:2010,Ryu:2014fo}, gas sensors~\cite{Schedin:2007,Wehling:2009hp,Nagarajan:2018eo}, very good thermal conductors~\cite{Wang:2018jz,Wu:2017bo}, etc.

Recently, a new class of low-dimensional materials, namely transition metal phosphorus trichalcogenides received a lot of attention, because they are considered as a new class of 2D materials, which can find application in electronics, sensing, catalysis, or energy conversion. Many experimental and theoretical works as well as several respective review articles on the studies of the transition metal phosphorus trichalcogenides appeared in literature~\cite{Li:2013ei,Li:2014de,Du:2016ft,Zhang:2016kra,Gusmao:2017kza,Susner:2017in,Wang:2018dha}. However, the available in literature experimental and theoretical works on the studies of, e.\,g. electrocatalytic properties of these materials, contain sometimes misleading information on the synthesis and characterization. These inaccurate results obtained during experiment/sample preparation and treatment of experimental data lead to the fact that the discussed effects and main conclusions are not supported by the presented experimental data. 

Recently a work of Gusm\~{a}o et al.~\cite{Gusmao:2017kza} was published, which is devoted to the studies of the electrochemical performance (hydrogen and oxygen evolution reactions) of transition metal phosphorus trichalcogenides MPSe$_3$ (M = Cd, Cr, Fe, Mn, Sn, Zn). It was found that among all synthesised samples, FePSe$_3$ and MnPSe$_3$ have the highest efficiency for the hydrogen evolution reaction with good stability. A the same time it was found that MnPSe$_3$ holds the lowest oxidation potential, although it was assigned to the presence of MnO$_2$ in the sample. In the same work MPSe$_3$ samples were synthesised using standard chemical vapour transport method. Structural characterization was performed by means of x-ray diffraction (XRD) in the $\theta-2\theta$ geometry using Cu\,K$\alpha$ line. Morphology and chemical analysis were studied using scanning electron microscopy (SEM) combined with energy-dispersive x-ray spectroscopy (EDS) applying the gentle electron beam with the energy of $2$\,keV. X-ray photoelectron spectroscopy (XPS) was used for the surface analysis of the studied samples. These data obtained during samples characterization were used to support the observed effects in further experiments on the electrochemical performance of these materials (hydrogen and oxygen evolution reactions). However, as shown below, the presented claims in the discussed manuscript have to be reconsidered, because experimental as well as theoretical data on the sample characterization contain several flaws and errors and these results have to be fully reanalyzed. Therefore the main conclusions on the electrochemical performance of transition metal phosphorus trichalcogenides have to be reconsidered.

\section{Experimental and computational details}\label{Exp_Details}

\textit{Samples synthesis.} Manganese ($99.9$\%), phosphorus ($99.999$\%), and selenium ($99.999$\%) from Shanghai Macklin Biochemical Co., Ltd. and Alfa Aesar were used during synthesis. A stoichiometric amount of high-purity elements (mole ratio $\mathrm{Mn}:\mathrm{P}:\mathrm{Se} = 1:1:3$, $1$\,g in total) and iodine (about $20$\,mg) as a transport agent were sealed into a quartz ampule (length $17$\,cm, external diameter approximately $15$\,mm) and kept in a two-zone furnace ($650-600^\circ$\,C). The pressure inside the ampule was pumped down to $1\times10^{-3}$\,Torr. After 10 days of heating, the ampule was cooled down to room temperature with bulk crystals in the colder edge.

\textit{Characterization.} XRD patterns were collected with a Bruker D2 Phaser diffractometer using Cu\,K$\alpha$ ($1.54178$\,\AA) radiation at room temperature. Optical images were collected with optical microscope at different magnifications. Presented reference spectra of WSe$_2$ were collected using SPECS PHOIBOS 150 energy analyzer and Al\,K$\alpha$ monochromatized x-ray source ($h\nu = 1486.6$\,eV).

\textit{DFT calculations.} Spin-polarised DFT calculations based on plane-wave basis sets of $500$\,eV cutoff energy were performed with the Vienna \textit{ab initio} simulation package (VASP)~\cite{Kresse:1996kg,Kresse:1994cp}. The Perdew-Burke-Ernzerhof (PBE) exchange-correlation functional~\cite{Perdew:1996} was employed. The electron-ion interaction was described within the projector augmented wave (PAW) method~\cite{Blochl:1994fq} with Mn ($3p$, $3d$, $4s$), Cr ($3p$, $3d$, $4s$), Fe ($3p$, $3d$, $4s$), P ($3s$, $3p$), and Se ($4s$, $4p$) states treated as valence states. The Brillouin-zone integration was performed on $\Gamma$-centred symmetry reduced Monkhorst-Pack meshes using a Gaussian smearing with $\sigma = 0.1$\,eV, except for the calculation of density of states (DOS). For these calculations, the tetrahedron method with Bl\"ochl corrections~\cite{Blochl:1994ip} was employed. A $12\times 12\times 4$ $k$-mesh was used in the case of ionic relaxations and $24\times 24\times 8$ for the DOS calculations, respectively. The DFT+$\,U$ scheme~\cite{Anisimov:1997ep,Dudarev:1998dq} was adopted for the treatment of Mn-, Cr- and Fe-$3d$ orbitals. Dispersion interactions were considered adding a $1/r^6$ atom-atom term as parameterised by Grimme (``D2'' parameterisation)~\cite{Grimme:2006fc}. During structure optimisation, the convergence criteria for energy and force were set equal to $10^{-5}$\,eV and $1\times10^{-2}$\,eV/\AA, respectively. 

\section{Results and discussions}\label{Results_Discussions}

\begin{figure}[h]
\centering
\includegraphics[width=0.75\textwidth]{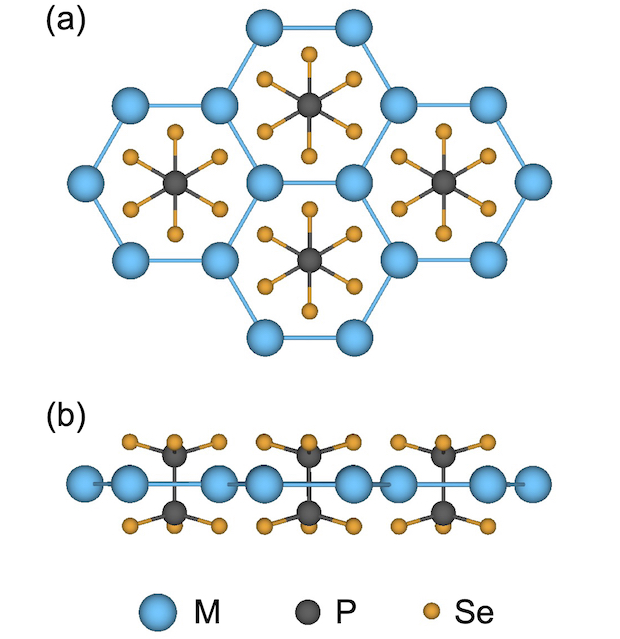}
\caption{Top (a) and side (b) views of a single layer of MPSe$_3$. Spheres of different size/colour represent ions of different type.}
\label{fig:structure}
\end{figure}

\begin{figure}[h]
\centering
\includegraphics[width=1.0\textwidth]{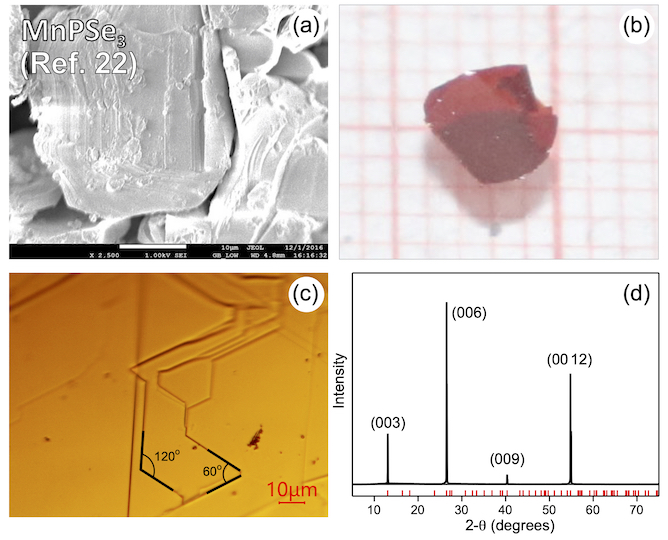}
\caption{(a) SEM image of MnPSe$_3$ extracted from Ref.~\citenum{Gusmao:2017kza} (white bar corresponds to $10\,\mu\mathrm{m}$). The MnPSe$_3$ crystals obtained in our experiments: (b) general view of the crystal (grid size is $1\,\mathrm{mm}\times1\,\mathrm{mm}$), (c) optical microscopy image, (d) respective XRD patterns.}
\label{fig:SEM_OPT_XRD}
\end{figure}

The studied 3D MPSe$_3$ crystals have either $C2/m$ (M = Cr) or $R\bar{3}$ (M = Mn, Zn, Cd, Fe) space group symmetry. The both cases can be viewed as layered structures, where each stacked MPSe$_3$ layer has a $D_{3d}$ symmetry (see Fig.~\ref{fig:structure}). Therefore, one might expect the hexagonal-like shape for the grown crystals with angles between crystallite edges either $120^\circ$ or $60^\circ$~\cite{Du:2016ft}. Analysis of the presented in Ref.~\citenum{Gusmao:2017kza} SEM images of the grown crystals demonstrate the absence of this representative feature, indicating the low quality of the studied crystals (see Fig.~\ref{fig:SEM_OPT_XRD}(a) for MnPSe$_3$; see also Figs.\,2 and S1 in Ref.~\citenum{Gusmao:2017kza} for other MPSe$_3$ crystals). At the same time, we present in Fig.~\ref{fig:SEM_OPT_XRD}(b,c) optical images of MnPSe$_3$ crystal grown recently in our laboratory. One can clearly see that this sample demonstrates very high quality with well ordered planes and steps oriented with respect to each other either by $120^\circ$ or $60^\circ$. The measured XRD patterns of our crystals demonstrate extraordinary quality without additional phases (Fig.~\ref{fig:SEM_OPT_XRD}(d))~\cite{Du:2016ft}, whereas the data presented in Ref.~\citenum{Gusmao:2017kza} (Fig.\,S3) demonstrate the existence of other undesired crystal phases in the studied samples.

The SEM/EDS combination was used in Ref.~\citenum{Gusmao:2017kza} to study the morphology as well as the composition of the studied MPSe$_3$ samples. Despite the low energy of the used electron beam during this analysis, the EDS method cannot be considered as a surface sensitive as, for example, at $5$\,keV of beam energy the probing depth can reach  $0.5\,\mu\mathrm{m}$. Therefore, the results presented in Fig.\,S2 and Tab\,S1 of Ref.~\citenum{Gusmao:2017kza} demonstrate extremely poor quality of the studied samples in bulk as well as at the surface: some samples demonstrate the absence of phosphorus in the sample and the level of the C- and O-contamination is very high (it is varied between $\approx43$\,at.\% and $\approx65$\,at.\% in total).

\begin{figure}[h]
\centering
\includegraphics[width=0.6\textwidth]{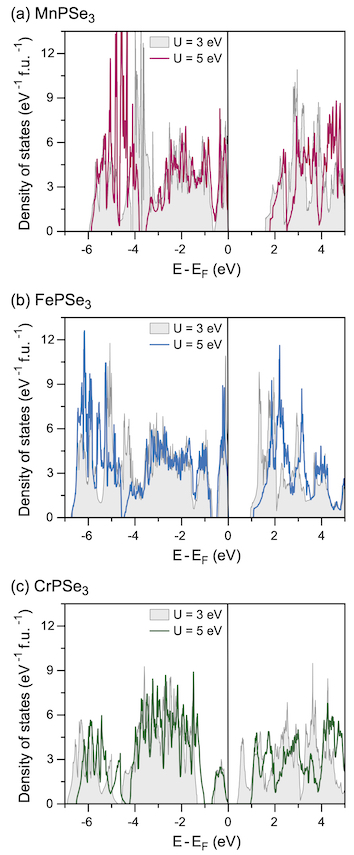}
\caption{DOS of bulk MPSe$_3$ (M = Mn, Fe, Cr) in the antiferromagnetic state calculated in the present work for $U=3$\,eV and $U=5$\,eV.}
\label{fig:MPX3_DOSs}
\end{figure}

The electronic structure of the studied MPSe$_3$ crystals was investigated in Ref.~\citenum{Gusmao:2017kza} using density functional theory (DFT) within the PBE+$U$ approach with $U=3$\,eV for the proper treatment of electron correlations in the valence band. Here we would like to point out, that from the presented in Ref.~\citenum{Gusmao:2017kza} data it is absolutely unclear if bulk 3D or monolayer 2D phases were considered in the DFT calculations. This is crucial for the description of the band gap around the Fermi level as its value strongly depends on the system dimensionality as shown in Ref.~\citenum{Yang:2020ex}. Thus, only comparison of experimental results with theoretical data obtained for the same phase (here, bulk 3D) will make sense. The calculated values for the band gap are summarized in Table~\ref{tab:band_gaps}, where we compare data from Ref.~\citenum{Gusmao:2017kza} and our own data obtained in the framework of the present study. Fig.~\ref{fig:MPX3_DOSs} shows density of states (DOS) plots for a series of MPSe$_3$ compounds calculated in the present study for the 3D phase and different values of $U$.

As can be seen, the values of band gap for MPSe$_3$ in Ref.~\citenum{Gusmao:2017kza} are approximately by factor of $2$ smaller compared to the experimental values, which appeared in earlier publications~\cite{Du:2016ft} and to the values obtained with DFT in the present study. These values are also not consistent with the recently published values for MnPSe$_3$~\cite{Yang:2020ex}, which are in good agreement with available experimental data. It is also very confusing, that in Ref.~\citenum{Gusmao:2017kza}, the band gap of $0.5$\,eV is claimed for FePSe$_3$, that contradicts to the metallic state for this compound as can be seen in Fig.\,3 of this reference. Moreover, high density of states at the Fermi level indicates an unstable situation, which is not energetically favourable.

When looking at the DFT results, i.\,e. DOS calculated for MPSe$_3$, one can note a significant variation of a band gap depending on the nature of M. Comparison of the calculated band gap with, e.g., the free energy of water splitting ($1.23$\,eV) may be used as an indication that MnPSe$_3$ (with $\Delta E_g=1.8$\,eV) is a good candidate for water-splitting catalyst. Such a consideration, however, is oversimplified, because for the strait forward conclusion one has to take into account some important factors. First of all, any catalytic process with MPX$_3$ will take place at the surface and it is known, that the electronic properties of these compounds strongly depend on the system dimensionality~\cite{Yang:2020ex}. Secondly, one has to consider presence of defects, which on the one hand are expected to enhance the catalytic activity~\cite{Li:2019cz}, on the other hand will influence (reduce) the band gap~\cite{Yang:2020ex}. In the case of electrochemical water splitting, the band edges of the catalysts must straddle the redox potentials of water, which in their turn depend on the $pH$ value~\cite{Zhang:2016kra}. Thus, in order to discuss the activity of MPSe$_3$ for the electrochemical applications, one has to perform very detailed, accurate, and systematic study, where the calculations on 3D MPSe$_3$ are just the initial step. At the same timeIt is necessary to note, that the scientific weight of the drawn conclusions will depend on the reliability of this initial step descriptions. 

The most confusing part of Ref.~\citenum{Gusmao:2017kza} regards to the XPS characterization of the studied MPSe$_3$ samples and their interpretation. Because these XPS data are used to support the respective electrochemical performance results for the studied material, we believe that the main conclusions of the discussed manuscript have to be reconsidered and the respective data have to be fully reanalized.

\begin{figure}[h]
\centering
\includegraphics[width=0.8\textwidth]{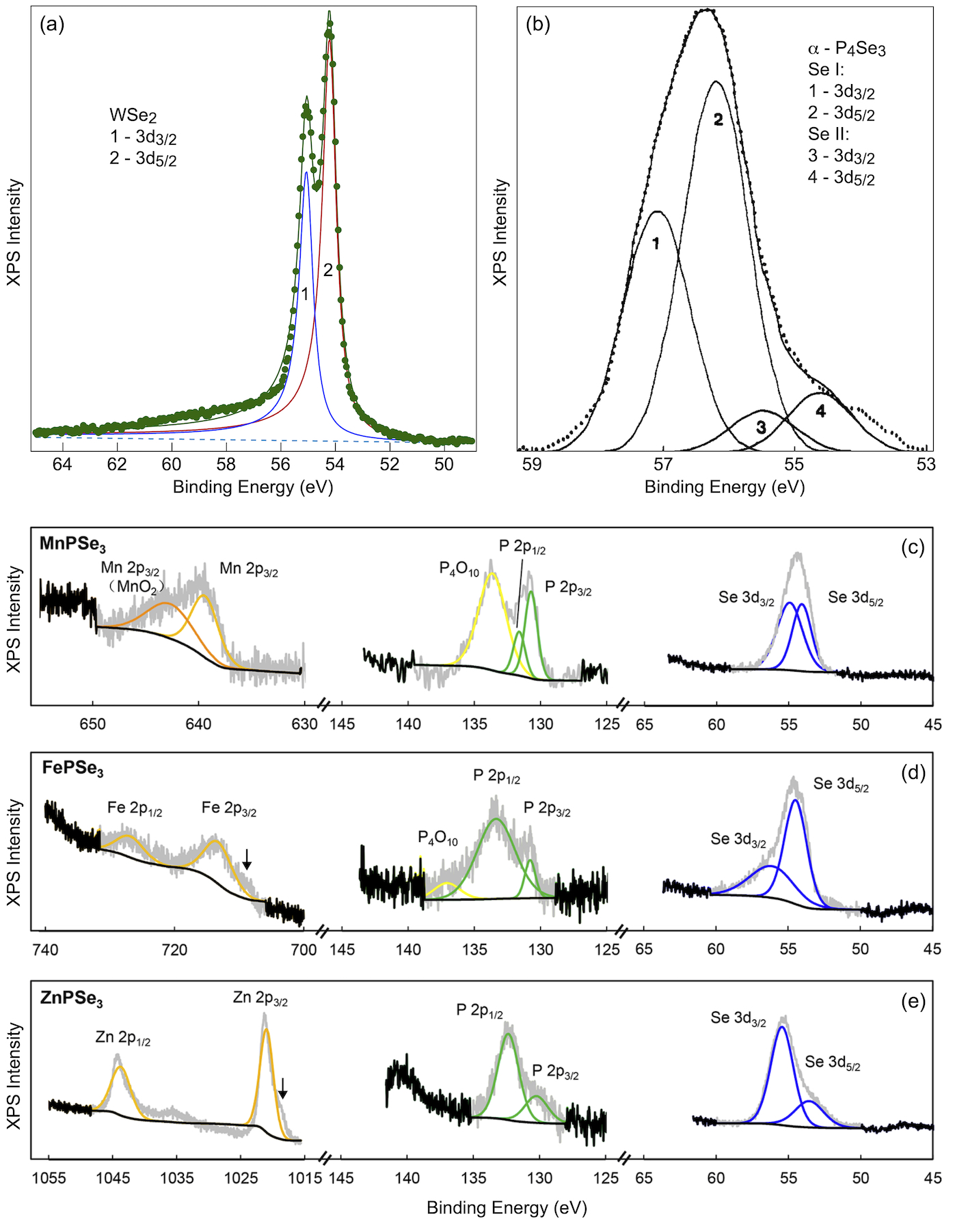}
\caption{(a) The Se\,$3d$ XPS spectrum measured for WSe$_2$ using monochromatized Al\,K$\alpha$ line. Respective spin-orbit split $3d_{3/2}$ and $3d_{5/2}$ emission lines are marked. (b) The Se\,$3d$ XPS spectrum for $\alpha$-P$_4$Se$_3$ measured using non-monochromatized Mg\,K$\alpha$ x-ray source~\cite{Rollo:1990ix}. Respective spin-orbit split emission lines for two kinds of P species are marked. (c-e) Examples of XPS spectra for MnPSe$_3$, FePSe$_3$, and ZnPSe$_3$ extracted from Ref.~\citenum{Gusmao:2017kza} and discussed in the present commentary.}
\label{fig:XPS}
\end{figure}

To remind, here we present the basic consideration for the analysis of XPS emission lines. Core levels in XPS use the notation $nl_j$, where $n$ is the principal quantum number, $l$ is the angular momentum quantum number and $j = l + s$ (where $s=\pm1/2$ is the spin angular momentum number). All orbital levels, except the $s$ levels ($l = 0$), give rise to a doublet with the two possible states having different binding energies. This is known as spin-orbit splitting. The peaks will also have specific area ratios based on the degeneracy of each spin state, i.e. the number of different spin combinations that can give rise to the total $j$. For example, for the $2p$ spectra, where $n=2$ and $l=1$, $j$ will be $1/2$ and $3/2$. The area ratio for the two spin orbit peaks ($2p_{1/2} : 2p_{3/2}$) will be $1 : 2$ (corresponding to $2$ electrons in the $2p_{1/2}$ level and $4$ electrons in the $2p_{3/2}$ level). The similar consideration is valid for $d$ and $f$ levels, where the area ratios for the respective spin-orbit split components are $2 : 3$ and $3 : 4$, respectively. These ratios must be taken into account when analyzing spectra of the $p$, $d$ and $f$ core levels. Spin-orbit splitting values (in eV) can be found in different databases (e.\,g.: https://xpssimplified.com/periodictable.php). Here, we present two representative examples, demonstrating the application of the fit procedure for the Se\,$3d$ spectra, where the above described peaks' ratios are used. The Se\,$3d$ spectrum for a WSe$_2$ single crystal measured using monochromotized Al\,K$\alpha$ x-ray source and shown in Fig.~\ref{fig:XPS}(a) demonstrates clear spin-orbit split doublet with components having equal full width at half maximum (FWHM), peak separation of $0.85$\,eV and intensities ratio of $I(3d_{3/2})/I(3d_{5/2})=0.67$. The Se\,$3d$ spectrum for $\alpha$-P$_4$Se$_3$ measured using non-monochromatized Mg\,K$\alpha$ x-ray source shows that all spin-orbit doublets must be fit in order to properly identify the species present in the sample. The $3d_{3/2}$ and $3d_{5/2}$ doublet for each chemical specie is constrained to have $\approx2 : 3$ peak area ratios, equal FWHM, and a peak separation of $\approx1$\,eV (Fig.~\ref{fig:XPS}(b))~\cite{Rollo:1990ix}. The discussed examples demonstrate the universality of the discussed approach without any connection to the specific sample.

Coming back to the XPS data presented in Ref.~\citenum{Gusmao:2017kza} we have to mention the high level of the C- and O-contamination that does not allow to carefully study the chemical states of elements in the studied compounds (see Fig.\,S5 of the discussed manuscript). Therefore, it is very difficult to draw the clear conclusions, which might support the further data obtained in the studies of the electrochemical performance of these materials. Moreover, the performed analysis and interpretation of the respective core levels presented in this manuscript does not correspond to the high standards accepted in the scientific publications. As an example, we present here several XPS spectra for MnPSe$_3$, FePSe$_3$, and ZnPSe$_3$ extracted from Fig.\,5 of Ref.~\citenum{Gusmao:2017kza}. The following list marks the serious deficiencies in the spectra interpretation (Fig.~\ref{fig:XPS}(c-e)):

\begin{itemize}

\item The respective fit of the Se\,$3d$ XPS lines is not correct for all data presented in the discussed manuscript. As discussed before, the ratio of integral intensities for $3d_{3/2}$ and $3d_{5/2}$ lines have to be $\approx0.67$ and cannot be more than $1$ as it is presented for the case of ZnPSe$_3$. If necessary, several spin-orbit split lines corresponding to different species have to be introduced for the proper fit of the XPS spectra.

\item The respective fit of P\,$2p$ XPS line is not correct for several data sets presented in the discussed manuscript (see the respective spectra for FePSe$_3$ and ZnPSe$_3$). As discussed before, the ratio of integral intensities for $2p_{1/2}$ and $2p_{3/2}$ lines have to be $\approx0.5$ and cannot be more than $1$ as it is presented for the case of ZnPSe$_3$. If necessary, several spin-orbit split lines corresponding to different species have to be introduced. It has to be also noted that the same broad peak in the P\,$2p$ spectra for different MPSe$_3$ compounds, which is located at $\approx134$\,eV of binding energy, is assigned to different emission lines: P$_4$O$_{10}$ or P\,$2p_{1/2}$, although it is obvious that this component corresponds to the same chemical state of P atoms, namely to phosphorus oxide (P$_x$O$_y$). Ironically, on the basis of this inaccurate fit, authors of Ref.~\citenum{Gusmao:2017kza} made the conclusion on the low concentration of P$_x$O$_y$ for FePSe$_3$ and on the absence of the oxidation state of P for ZnPSe$_3$, which is, of course, wrong.

\item For the Mn\,$2p_{3/2}$ XPS line, two components in the spectra were assigned to the emission from Mn atoms either in MnPSe$_3$ or in MnO$_2$ (Fig.~\ref{fig:XPS}(c)). However, this statement is not supported by any additional reference XPS measurements. The quality of the XPS spectra presented in Ref.~\citenum{Gusmao:2017kza} is very poor, not allowing to perform clear identification of the components and several Mn-oxide phases can be assigned to this ``MnO$_2$''-peak: MnO, Mn$_2$O$_3$, or MnO$_2$~\cite{Biesinger:2011de}.

\item The presented fit of the Fe\,$2p$ XPS spectra is not correct. One can clearly see a well resolved shoulder at the lower binding energies for Fe\,$2p_{3/2}$ (marked by the arrow in Fig.~\ref{fig:XPS}(d)). This component is missed in the analysis; however, it might be assigned to another chemical state of Fe atoms in FePSe$_3$, or the main fitted peak can be due to Fe$_x$O$_y$ in the studied sample~\cite{Dedkov:2008e,Dedkov:2011cm} and the small not-assigned peak is due to the FePSe$_3$ phase.

\item The previous consideration is also valid for the analysis of the Zn\,$2p$ XPS spectra. One can clearly see a well resolved shoulder at the lower binding energies for Zn\,$2p_{3/2}$ (marked by the arrow in Fig.~\ref{fig:XPS}(e)), which is also is not included in the respective analysis~\cite{Wang:2015bv,Kundu:2016gs}.

\item Analysis of Table\,S2 of Ref.~\citenum{Gusmao:2017kza} shows that binding energies for the spin-orbit split components of the Se\,$3d$ XPS line change the order for different compounds, which is, of course, wrong - the binding energy of the $3d_{5/2}$ has to be always smaller. However, this mistake can be related to the typing error.

\end{itemize}

\section{Conclusions}

In conclusion, we have demonstrated that the recent work of Gusm\~{a}o et al.~\cite{Gusmao:2017kza} contains several serious flaws and errors, which are related to the characterization of the synthesised MPSe$_3$ samples (M = Cd, Cr, Fe, Mn, Sn, Zn). In this work the structural characterization was performed using XRD and sample morphology and chemical analysis was studied using the SEM/EDS combinations. XPS was used for the surface analysis of the studied samples. Because these data are considered as the main results for the MPSe$_3$ samples characterization and they are used to support the respective results from the further studies of the electrochemical performance of the studied objects (hydrogen and oxygen evolution reactions), we believe that the main conclusions of the discussed manuscript are not valid and the respective experimental data, which are criticised here, have to be carefully reconsidered and reanalyzed. Here we would like to point out that the recently published data have no limitation period because all presently used experimental and theoretical approaches were and are available for the last 10 years (at least), as demonstrated by the respective examples. 

\section*{Acknowledgement}

This contribution was supported by the National Natural Science Foundation of China (Grant No. 21973059). Y.D. and E.V. thank the support by the Ministry of Science and Higher Education of the Russian Federation (State assignment in the field of scientific activity, Southern Federal University, 2020).

\section*{References}


\clearpage
\begin{table}
\caption{Comparison of band gaps obtained in different works using theoretical (DFT) and experimental methods.}
\label{tab:band_gaps}
\renewcommand{\arraystretch}{1.2}
\begin{tabular}{l l lll l l }
\hline
Compound & PBE+$U$ & \multicolumn{3}{c}{PBE+$U$+D2} & HSE06+D2 & Experiment \\
		  & $U$=\,3\,eV (Ref.~\citenum{Gusmao:2017kza})& $U$=\,3\,eV & $U$=\,4\,eV & $U$=\,5\,eV &(Ref.~\citenum{Yang:2020ex}) &(Ref.~\citenum{Du:2016ft}) \\
\hline
MnPSe$_3$	&$1.40$\,eV&$1.60$\,eV	&$1.72$\,eV	&$1.80$\,eV	&$2.50$\,eV	&$2.25$\,eV\\
SnPSe$_3$	&$1.20$\,eV&			&			&			&			&	\\	
CrPSe$_3$	&$0.25$\,eV&$0.39$\,eV	&$0.70$\,eV	&$0.96$\,eV	&			&	\\
FePSe$_3$	&$0.50$\,eV&$0.95$\,eV	&$1.04$\,eV	&$1.08$\,eV	&			&$1.20$\,eV	\\
\hline
\end{tabular}
\end{table}

\clearpage

\clearpage

\clearpage

\clearpage

\end{document}